\documentstyle[aps,eqsecnum,preprint]{revtex}

\newcommand{\sgn}{\rm sgn}

\begin{document}
\draft
%\tighten                                           %   Causes single spacing

\title{Qualitative Analysis of Causal Anisotropic Viscous Fluid Cosmological Models}
\author{R.J. van den Hoogen  and A.A. Coley }
\address{Department of Mathematics, Statistics, and Computing Science, Dalhousie University, Halifax, Nova Scotia, Canada, B3H 3J5}
\date{\today}
\maketitle

\begin{abstract}

The truncated Israel-Stewart theory of irreversible thermodynamics is used to describe the bulk viscous pressure and the anisotropic stress in a class of spatially homogeneous viscous fluid cosmological models.  The governing system of differential equations is written in terms of dimensionless variables and a set of dimensionless equations of state is utilized to complete the system. The resulting dynamical system is then analyzed using standard geometric techniques.
It is found that the presence of anisotropic stress plays a dominant role in the evolution of the anisotropic models. In particular, in the case of the Bianchi type I models it is found that anisotropic stress leads to models that violate the weak energy condition and to the creation of a periodic orbit in some instances.  The stability of the isotropic singular points is analyzed  in the case with zero heat conduction;  it is found that there are ranges of parameter values such that there exists an attracting isotropic Friedmann-Robertson-Walker model.  In the case of zero anisotropic stress but with non-zero heat conduction the stability of the singular points is found to be the same as in the corresponding case with zero heat conduction; hence the presence of heat conduction does not apparently affect the global dynamics of the model.  
 
\end{abstract}

\medskip

%0000000000000000000000000000000000000000000000000000 
%000000000000          INTRODUCTION      000000000000 
%0000000000000000000000000000000000000000000000000000 

\section{Introduction}\label{I}

Recently spatially homogeneous and isotropic imperfect fluid cosmological models
 were investigated using techniques from dynamical systems theory \cite{Coley95}.  In these imperfect fluid models the bulk viscous pressure satisfies the truncated Israel-Stewart theory of irreversible thermodynamics.  However, the models studied in \cite{Coley95} are isotropic Friedmann-Robertson-Walker (FRW) models which do not allow processes such as shear viscous stress; consequently the next step in this research programme is to study  anisotropic generalizations of the FRW models.  It was argued in \cite{Coley95} that the  anisotropic models studied in \cite{Coley94a,Abolghasem93,Coley92,Burd94} that allow shear viscous stress and which satisfy the Eckart theory of irreversible thermodynamics are not satisfactory since
viscous signals in the fluid could travel faster than the speed of light -- therefore it was concluded that   a causal theory of irreversible thermodynamics such as the truncated Israel-Stewart theory ought to be utilized.

Assuming that the universe can be modelled as a simple fluid and omitting certain divergence terms, the truncated Israel-Stewart equations for the bulk viscous pressure, $\Pi$, the heat conduction vector, $q_a$, and the anisotropic stress, $\pi_{ab}$ (or shear viscous stress), are given by \cite{IsraelStewart79}:
\begin{mathletters}\label{israel}
\begin{eqnarray}
\Pi &=& -\zeta(u^a_{\:;a}+\beta_0\dot\Pi-\alpha_0q^a_{\:;a}), \label{israel1}\\
q^a &=& -\kappa T h^{ab}(T^{-1}T_{;b}+\dot u_b+\beta_1\dot q_b-\alpha_0 \Pi_{;b}-\alpha_1\pi^c_{\:b;c}),\label{israelq}\\
\pi_{ab}&=& -2\eta\langle u_{a;b} +\beta_2\dot\pi_{ab}-\alpha_1q_{a;b}\rangle,
\end{eqnarray}
\end{mathletters}
where $u_a$ is the fluid four-velocity, $h_{ab}=u_au_b+g_{ab}$ is the projection tensor and 
$\langle A_{ab}\rangle\equiv \frac{1}{2}h^c_{\:a}h^d_{\:b}(A_{cd}+A_{dc}-\frac{2}{3}h_{cd}h^{ef}A_{ef})$.
The variable $T$ represents the temperature, $\kappa$ represents the thermal conductivity, $\beta_0$, $\beta_1$ and $\beta_2$ are proportional to the relaxation times, $\alpha_0$ is a coupling parameter between the heat conduction and the bulk viscous pressure, and $\alpha_1$ is a coupling parameter between the anisotropic stress and the heat conduction.
We shall refer to equations (\ref{israel}) as the truncated Israel-Stewart equations.  Equations (\ref{israel})  reduce to the Eckart equations  used in \cite{Coley94a,Abolghasem93,Coley92,Burd94} when $\alpha_0=\alpha_1=\beta_0=\beta_1=\beta_2=0$.

Belinskii et al. \cite{Belinskii80} were the first to study  cosmological models using the truncated Israel-Stewart theory of irreversible thermodynamics to model the bulk viscous pressure and the anisotropic stress.   Using qualitative analysis, Bianchi type I models were investigated subject to equations
 (\ref{israel}) in which equations of state of the form 
\begin{equation}
\zeta=\zeta_0\rho^{m}, \quad \eta=\eta_0\rho^{n}, \quad \beta_0 = \rho^{-1}, 
\quad  \text {and} \quad \beta_2 = \rho^{-1}, \label{Bel eqs state}
\end{equation}
were assumed \cite{Belinskii80},
where $m$ and $n$ are constants and $\zeta_o$ and $\eta_o$ are parameters.
The isotropizing effect found in the Eckart models no longer necessarily occurred in the truncated Israel-Stewart models.
It was also found that the cosmological singularity still exists but is of a new type, namely one with an accumulated  ``visco-elastic'' energy \cite{Belinskii80}.

Recently, Romano and Pav\'on \cite{Romano93,Romano94} have studied anisotropic cosmological models in a causal theory of irreversible thermodynamics, analyzing the stability of the isotropic singular points in the Bianchi type I and III models.  They also assumed equations of state of the form (\ref{Bel eqs state}). However, they concluded that any initial anisotropy dies away rapidly but the shear viscous stress need not vanish and hence neither the de Sitter models  nor the Friedmann models are attractors.

In this paper we shall  analyze qualitatively  a class of anisotropic cosmological models arising from the use of the truncated Israel-Stewart equations, thereby expanding the analysis in \cite{Coley95} to anisotropic
models and extending the analysis in \cite{Coley94a,Abolghasem93,Coley92,Burd94} to causal theories.  
We will analyze both the Bianchi type V and the Bianchi type I models, which are simple generalizations of the open and flat FRW models.  The system of ordinary differential equations governing the models is a dynamical system.  We will find the singular points and determine their stability.  
In previous work  \cite{Coley95,Coley94a,Abolghasem93,Coley92,Burd94} dimensionless variables and a set of dimensionless equations of state were employed to analyze various spatially homogeneous imperfect fluid  cosmological models.  We shall utilize the same dimensionless variables and dimensionless equations of state here.
One reason for using dimensionless equations of state is that the equilibrium points of the system of differential equations describing the spatially homogeneous models will represent self-similar cosmological models \cite{Coley94b}.
In addition, it could be argued that the use of dimensionless equations of state is   natural  at least in some physical situations of interest (see e.g. \cite{Coley90b}).  (See also Coley \cite{Coley90a}, and Coley and van den Hoogen \cite{Coley94a} for further motivation for using these dimensionless equations of state.)

In section~\ref{II} we define the models and establish the resulting dynamical system.
 In section~\ref{III}  we investigate the qualitative behaviour of the system for different values of the physical parameters.  In particular, in section~\ref{III.1} we analyze the system when there is no heat conduction and in section~\ref{III.3} we analyze the system when there is heat conduction and bulk viscosity but with no anisotropic stress.    In section~\ref{IV} we  discuss  our results and make a few concluding remarks. 
 For simplicity we have chosen units in which $8\pi G = c= 1$.

%0000000000000000000000000000000000000000000000000000 
%000000000000             Bianchi V      000000000000 
%0000000000000000000000000000000000000000000000000000 

\section{Anisotropic Models}\label{II}

The Bianchi type V models are anisotropic   generalizations of the negative curvature Friedmann-Robertson-Walker (FRW) cosmological models.
The diagonal form of the Bianchi type V metric is given by: 
\begin{equation}
ds^2 = -dt^2 + a(t)^2dx^2 + b(t)^2e^{2x}dy^2 +
c(t)^2e^{2x}dz^2 .\label{gen.metric}
\end{equation}
We assume that the fluid is moving orthogonal to the homogeneous spatial hypersurfaces; that is, the fluid 4-velocity, $u^a$, is equal to the unit normal of the spatial hypersurfaces.
The energy-momentum tensor can be decomposed with respect to $u^a$ according to \cite{MacCallum73}:
\begin{equation}
T_{ab}=(\rho+\bar p)u_au_b+\bar p g_{ab} + q_au_b + u_aq_b+\pi_{ab},\label{emtensor}
\end{equation}
where $\bar p \equiv p + \Pi$ and 
$\rho$ is the energy density, 
$p$ is  the thermodynamic pressure,  
$\Pi$ is the bulk viscous pressure, 
$\pi_{ab}$ is the anisotropic stress, and 
$q_a$ is the heat conduction vector as measured by an observer moving with the fluid.

The Einstein field equations and the energy conservation equations  in terms of the expansion($\theta$) and shear($\sigma$) are: (see Coley and van den Hoogen \cite{Coley94a}):
\begin{mathletters}\label{2.3}
\begin{eqnarray}
\dot\theta &=& - 2\sigma^2 - {{1}\over3}\theta^2 - {{1}\over2}(\rho + 3p+3\Pi),\label{raychaudhuri}\\
\dot\rho  &=&   - \theta(\rho + p +\Pi) - {{2}\over {a^2}}q_1-\frac{1}{3}\biggl(\sigma_1(2\Pi_1-\Pi_2)+\sigma_2(2\Pi_2-\Pi_1)\biggr), \label{rho} \\
\dot\sigma_1 &=&   - \theta\sigma_1+\Pi_1,\label{sigma1}\\
\dot\sigma_2 &=&   - \theta\sigma_2+\Pi_2,\label{sigma2}
 \end{eqnarray}
\end{mathletters}
\begin{mathletters}\label{2.4}
\begin{eqnarray}
\theta^2&=&  3\sigma^2 +3\rho  + {{9}\over {a^2}}, \label{gen.equality}\\
q_1 &=& -\sigma_1 -\sigma_2,\label{def.q}
 \end{eqnarray} 
\end{mathletters}
where $\sigma^2=\frac{1}{3}(\sigma_1+\sigma_2)^2 - \sigma_1\sigma_2$.  We have used the property that both $\sigma^a_{\ a}=0$ and $\pi^a_{\ a}=0$ to define new shear variables $\sigma_1=\sigma^1_{\ 1}-\sigma^2_{\ 2}$ and $\sigma_2=\sigma^1_{\ 1}-\sigma^3_{\ 3}$ and new anisotropic stress variables $\Pi_1=\pi^1_{\ 1}-\pi^2_{\ 2}$ and $\Pi_2=\pi^1_{\ 1}-\pi^3_{\ 3}$ to simplify the system.

The evolution equations for $\Pi$, $\Pi_1$, and $\Pi_2$ in this particular model, derived from (\ref{israel}), and using (\ref{2.4}), are then given by:
\begin{mathletters}\label{newpi_s}
\begin{eqnarray}
\dot\Pi&=&-\frac{\Pi}{\beta_0\zeta}-\frac{1}{\beta_0}\left(\theta+\frac{2}{9}\alpha_0(\sigma_1+\sigma_2)(\theta^2-3\sigma^2-3\rho)\right),\label{newpi}\\
\dot\Pi_1&=&-\frac{\Pi_1}{2\eta\beta_2}-\frac{1}{\beta_2}\left(\sigma_1-\frac{1}{9}\alpha_1(\sigma_1+\sigma_2)(\theta^2-3\sigma^2-3\rho)\right),\label{newpi1}\\
\dot\Pi_2&=&-\frac{\Pi_2}{2\eta\beta_2}-\frac{1}{\beta_2}\left(\sigma_2-\frac{1}{9}\alpha_1(\sigma_1+\sigma_2)(\theta^2-3\sigma^2-3\rho)\right).\label{newpi2}
\end{eqnarray}
\end{mathletters}
Note that the heat conduction $q_1$ is completely determined by the shear via equation (\ref{def.q}); thus equation (\ref{israelq}) is not needed for the  determination of the asymptotic behaviour of the models.

Now the system of equations  (\ref{2.3}) and (\ref{newpi_s}), define a dynamical system for the quantities $(\theta,\rho, \sigma_1, \sigma_2,\Pi, \Pi_1, \Pi_2)={\bf X}$ of the form $\dot {\bf X} = {\bf F}({\bf X})$.  This system of equations is invariant under the mapping (see Coley and van den Hoogen \cite{Coley94a,Coley94b})
\begin{equation}
\begin{array}{llll}
\theta \to\lambda\theta, & \sigma_1\to \lambda\sigma_1, &\sigma_2\to \lambda\sigma_2, & p\to\lambda^2 p, \\
\rho \to\lambda^2\rho,\qquad & \Pi \to \lambda^2 \Pi, \qquad & \Pi_1 \to \lambda^2\Pi_1, \qquad & \Pi_2 \to \lambda^2 \Pi_2, \\
\zeta \to \lambda \zeta, & \eta \to \lambda \eta, & \beta_0\to\lambda^{-2} \beta_0, &\beta_2 \to \lambda^{-2}\beta_2, \\
 \alpha_0 \to \lambda^{-2}\alpha_0,  & \alpha_1 \to \lambda^{-2}\alpha_1, & t\to{\lambda}^{-1}t,  &  
\end{array}\label{transformation}
\end{equation}
and this invariance implies that there exists a symmetry \cite{Bluman} in the dynamical system  and hence  a change of variables such that one of the equations can be made to decouple from the system.

We define new dimensionless variables $x$, $\Sigma_1$, $\Sigma_2$, $y$, $z_1$, $z_2$ and
a new time variable $\Omega$ as follows: 
$$
x \equiv {{3\rho}\over {\theta^2}}, \qquad
\Sigma_1 \equiv {{2\sqrt{3}\sigma_1}\over \theta}, \qquad
\Sigma_2 \equiv {{2\sqrt{3}\sigma_2}\over \theta}, \qquad
y\equiv {{9\Pi}\over{\theta^2}},
$$
\begin{equation}
 z_1\equiv\frac{\sqrt{3}\Pi_1}{2\theta^2}, \qquad
 z_2\equiv\frac{\sqrt{3}\Pi_2}{2\theta^2}, \quad{\text{ and }}\quad {{d\Omega}\over {dt}} = - {{1}\over 3}\theta.\label{new vars}
\end{equation}
With this  choice of variables, the Raychaudhuri equation,  (\ref{raychaudhuri}), effectively decouples from the system.

In order to complete the system of equations we need to specify
equations of state for the quantities $p$, $\zeta$, $\eta$, $\beta_0$, $\beta_2$, $\alpha_0$ and $\alpha_1$.  In principle,
equations of state can be derived from kinetic theory, but in practice one must
specify phenomenological equations of state which may or may not have any
physical foundations.   Following Coley
\cite{Coley90b,Coley90a}, we introduce dimensionless equations of state of the
form:
\begin{eqnarray}
{p\over {\theta^2}}                   = p_ox^{\ell}, && \nonumber\\
{{\zeta}\over {\theta}}               = \zeta_ox^m,  \qquad &&
{{\eta}\over {\theta}}                = \eta_ox^n,   \nonumber\\
{{3}\over {\beta_0\theta^2}}          = a_1x^{r_1},  \qquad &&
{{3}\over {4\beta_2\theta^2}}         = a_2x^{r_2}, \label{equations of state}
 \\
{{\alpha_0\theta^2}\over{4\sqrt{3}}}  = d_1x^{p_1},  \qquad &&
{{\alpha_1\theta^2}\over {36}}        = d_2x^{p_2},   
\nonumber\end{eqnarray}
where $p_o$, $\zeta_o$, $\eta_o$, $a_i$ and $d_i$ ($1\leq i \leq2$) are positive constants, and $\ell$, $m$, $n$, $r_i$ and $p_i$ ($1\leq i \leq2$) are constant parameters ($x$ is the
dimensionless density parameter defined earlier).  In the models under consideration
  $\theta$ is strictly positive, thus equations (\ref{equations of state}) are well defined.

We define new constants $b_1=a_1/\zeta_o$, $b_2=a_2/\eta_o$, $c_1=a_1d_1$ and $c_2=a_2d_2$. The system of equations  (\ref{2.3}),  and (\ref{newpi_s}),  written in the new dimensionless variables (\ref{new vars}) and employing the above equations of state (\ref{equations of state}), then  become
\begin{mathletters}
\label{sys of equations}
\begin{eqnarray}
\frac{dx}{d\Omega}&=&  x(1-2q)+9p_ox^l+y+\Sigma_1(2z_1-z_2)+\Sigma_2(2z_2-z_1)\nonumber\\
&& \qquad\qquad\qquad\qquad\qquad\qquad-\frac{1}{4\sqrt{3}}(\Sigma_1+\Sigma_2)(4-4x-\Sigma^2), \\
\frac{d\Sigma_1}{d\Omega} &=& \Sigma_1(2-q) -12z_1,\label{Sigma1}\\
\frac{d\Sigma_2}{d\Omega} &=& \Sigma_2(2-q) -12z_2,\label{Sigma2}\\
\frac{dy}{d\Omega}        &=& y(b_1x^{r_1-m}-2-2q)+ 9a_1x^{r_1}+c_1x^{p_1+r_1}(\Sigma_1+\Sigma_2)(4-4x-\Sigma^2),\\
\frac{dz_1}{d\Omega}      &=& z_1(2b_2x^{r_2-n}-2-2q) + a_2x^{r_2}\Sigma_1-c_2x^{p_2+r_2}(\Sigma_1+\Sigma_2)(4-4x-\Sigma^2),\\
\frac{dz_2}{d\Omega}      &=& z_2(2b_2x^{r_2-n}-2-2q) + a_2x^{r_2}\Sigma_2-c_2x^{p_2+r_2}(\Sigma_1+\Sigma_2)(4-4x-\Sigma^2),
\end{eqnarray}
\end{mathletters}
where $\Sigma^2\equiv\frac{1}{3}(\Sigma_1+\Sigma_2)^2-\Sigma_1\Sigma_2$. The quantity $q$ is the generalized dimensionless deceleration parameter given by
\begin{equation}
q\equiv\frac{-\ddot \ell \,\ell}{\dot \ell^2}=\frac{1}{2}\left(x+y+9p_ox^l+\Sigma^2\right),
\end{equation}
where $\ell$ is the average length scale of the universe (i.e., $\theta=3\frac{\dot\ell}{\ell}$).

Finally, from the Friedmann equation, (\ref{gen.equality}), we obtain the inequality
\begin{equation}
4-4x-\Sigma^2=\frac{36}{a^2\theta^2}\geq0
\end{equation}
where equality implies that the model is of Bianchi type I.  The interior of the parabola $4=\Sigma^2+4x$ in the phase space represents models  of Bianchi type V, while the parabola itself represents models of Bianchi type I.   There are other physical constraints that may be imposed, for example the energy conditions \cite{HawkingEllis}, which may place bounds on the variables $x$, $y$, $\Sigma_1$, $\Sigma_2$, $z_1$, and $z_2$.  A full list of the energy conditions is given in  Appendix~\ref{appendixA}.  In the present work we shall always assume that $x\geq0$, which states that the energy density in the rest frame of the matter is non-negative. This is a necessary condition for the fulfillment of   the weak energy condition (WEC) \cite{Kolassis88}.

The equilibrium points of the above system   all represent self-similar cosmological models, except for those singular points that satisfy $\dot\theta/\theta^2=-(q+1)/3=0$.    If $q \not = -1$, the nature of the equations of state (\ref{equations of state}) at the equilibrium points, is independent of the parameters $l$, $m$, $n$, $r_1$, $r_2$, $p_1$, and $p_2$, and  is given by
\begin{equation}
p \propto \rho, \qquad \zeta \propto \rho^{\frac{1}{2}}, \qquad \eta\propto \rho^{\frac{1}{2}}
\nonumber\end{equation}
\begin{equation}
\beta_0 \propto \rho^{-1}, \qquad \beta_2 \propto \rho^{-1},\qquad \alpha_0 \propto \rho^{-1},\quad \text{ and } \quad \alpha_1 \propto \rho^{-1}.
\end{equation}
Therefore  natural choices for $l$, $m$, $n$, $r_1$, $r_2$, $p_1$ and $p_2$ are respectively $1$, $1/2$, $1/2$, $1$, $1$, $-1$, $-1$. 
We note that if there exists a singular point with  $q=-1$, then it necessarily represents a de Sitter type solution which is not self-similar.

The most commonly used equation of state for the pressure  is the barotropic
equation of state $p = (\gamma - 1)\rho$, whence from (\ref{equations of state}) $p_o = {{1}\over 3} (\gamma -
1)$ and $l= 1$ (where $1 \leq \gamma\leq 2$ is necessary for local mechanical
stability and for the speed of sound in the fluid to be no greater than the speed of
light).  In addition, $l=1$ reflects the asymptotic behaviour of the equation of state for $p$.

To further motivate the choice  of the parameter ${r_1}$, we consider the velocity of a viscous pulse in the fluid \cite{Zakari93},
\begin{equation}
v=\left(\frac{1}{\rho\beta_0}\right)^{1/2},
\end{equation}
where $v=1$ corresponds to  the speed of light. Using (\ref{new vars}) and equations (\ref{equations of state}), we obtain
\begin{equation}
v=(a_1x^{{r_1}-1})^{1/2}.
\end{equation}
Now, if ${r_1}=1$, then not only do we obtain the correct asymptotic behaviour 
  of the equation of state for the quantity $\beta_0$ but we  also
   choose $0<a_1<1$ since then the velocity of a viscous pulse will be less than the velocity of light for any value of the density parameter $x$. 
In this way the parameter $a_1$ has a physical interpretation as the square of the speed of a viscous pulse in the fluid. 
Therefore, in the remainder of this analysis we shall choose ${r_1}=r_2=1$.  We shall  also choose $m=n=1$ and $p_1=p_2=-1$ for simplicity.  

Using these particular values for $m$, $n$, $r_1$, $r_2$, $p_1$, and $p_2$, we can easily show that all singular points are self-similar except in the case $\gamma =3\zeta_o$, whence the singular point $(x,\Sigma_1,\Sigma_2,y,z_1,z_2)= (1,0,0,-3\gamma,0,0)$ represents a de Sitter model. This is precisely the same as in the case  in which the Eckart theory was employed \cite{Coley94a}.

The full six-dimensional system is very difficult to analyze completely, so various physically interesting subsystems are investigated.  The case of zero heat conduction implies, via equations (\ref{def.q}) and (\ref{new vars}) that $\Sigma_1+\Sigma_2=0$.  In addition, adding equations (\ref{Sigma1}) and (\ref{Sigma2}) we   deduce that  $z_1+z_2=0$, in which case the resulting system is four-dimensional (see Section~\ref{III.1}).  
The case of non-zero heat conduction with zero anisotropic stress is a three dimensional system and is discussed in Section~\ref{III.3}.

%0000000000000000000000000000000000000000000000000000 
%00000000000000000000        SECTION  3         00000
%0000000000000000000000000000000000000000000000000000 

\section{Qualitative Analysis}\label{III}

The qualitative analysis of a system of autonomous ordinary differential equations $\dot{\bf X}  = {\bf F}({\bf X})$ begins with the determination of the singular points, that is, those points ${\bf X}$ such that ${\bf F}({\bf X})={\bf 0}$.  To analyze the stability of each singular point we linearize the system in a neighborhood of each singular point.  The signs of the eigenvalues of the derivative matrix determines the stability of the singular point, provided that the eigenvalues have non-zero real parts. (See Hirsch and Smale \cite{Hirsch}, Sansone and Conti \cite{Sansone}, Wiggins \cite{Wiggins} and Andronov et al. \cite{Andronov} for useful reviews.)  

\subsection{Zero Heat Conduction}\label{III.1}

In the case of zero heat conduction, $q_1=0$,   the field 
equations imply that $\Sigma_1+\Sigma_2=0$ and $z_1+z_2=0$.  Also,  
$\Sigma^2=\Sigma_1^{\ 2}$, hence (for simplicity), we shall drop the subscripts
on $\Sigma$ and $z$; that is, $\Sigma\equiv\Sigma_1=-\Sigma_2$ and $z\equiv z_1=-z_2$.
The system of equations then becomes:
\begin{mathletters}\label{zero heat eqs}
\begin{eqnarray}
x'&=&x(3\gamma-2-2q)+y+6z\Sigma,\\
\Sigma'&=&  \Sigma(2-q)-12z,\\
y'     &=&  y(b_1-2-2q)+9a_1 x,\\
z'     &=&  z(2b_2-2-2q)+a_2x\Sigma,
\end{eqnarray}
\end{mathletters}
where
\begin{equation}
q     =\frac{1}{2}\left((3\gamma-2)x+y+\Sigma^2\right),
\end{equation}
and the physical phase space is
\begin{equation}
4-4x-\Sigma^2\geq 0 \qquad\text{ and }\qquad x\geq 0.
\end{equation}

There exists three obvious and physically motivated invariant sets in
the phase space of the system.  They are ${\cal FRW}:=\{(x,\Sigma,y,z)\vert 
\Sigma=z=0\}$, ${\cal BI}:=\{(x,\Sigma,y,z) \vert 4-4x-\Sigma^2=0, \text{ and }  \Sigma \not = 0\}$, and ${\cal BV}:= {\cal BI}^c \cap {\cal FRW}^c$ (where subscript $c$ denotes the 
complement) which represents the Bianchi type V models.  
The set ${\cal FRW}$ represents the spatially homogeneous
and isotropic negative and flat curvature FRW models and the set ${\cal BI}$ represents the Bianchi type I models.  There are possibly eleven
different singular points of the system (\ref{zero heat eqs}).
The singular points lying in the set ${\cal FRW}$ are
\begin{equation}
(0,0,0,0),\qquad
(1,0,y^-,0),\qquad (1,0,y^+,0),
\end{equation}
where 
\begin{equation}
y^\pm=\frac{b_1-3\gamma}{2}\pm\frac{1}{2}\sqrt{(b_1-3\gamma)^2+36a_1}.\label{y pm}
\end{equation}
Also, if $B_1=0$ then there is a non-isolated line of singular points that passes through the points $(0,0,0,0)$ and $(1,0,y^-,0)$, where $B_1$ is,
\begin{equation}
B_1=(3\gamma-2)(2-b_1)+9a_1.
\label{B1}
\end{equation}
These points represent open ($x=0$) and flat ($x=1$) FRW models.  There are
possibly six singular points lying in the ${\cal BI}$ invariant set.  
They are
\begin{equation}
(0,-2,0,0), \qquad (0,+2,0,0)
\end{equation}
which represent Kasner models, and
\begin{eqnarray}
&&(\bar x^+,+\bar\Sigma^+,\bar y^+,+\bar z^+),\qquad(\bar x^+,-\bar\Sigma^+,\bar y^+,-\bar z^+)\nonumber\\
&&(\bar x^-,+\bar\Sigma^-,\bar y^-,+\bar z^-),\qquad(\bar x^-,-\bar\Sigma^-,\bar y^-,-\bar z^-)
\end{eqnarray}
where
\begin{eqnarray*}
\bar x^{\pm}&=&\frac{(\bar q^{\pm}-2)}{6a_2}(b_2-1-\bar q^{\pm}),\\
(\bar\Sigma^{\pm})^2&=&4-4\bar x^{\pm},\\
\bar y^{\pm}&=&\frac{(\bar q^{\pm}-2)}{2a_2}\biggl(4a_2+(2-\gamma)
(b_2-1-\bar q^{\pm})  
\biggr),\\
\bar z^{\pm} &=& \frac{-\bar\Sigma^{\pm}}{12}(\bar q^{\pm}-2),
\end{eqnarray*}
and
$\bar q^{\pm}$ is given by
\begin{equation}
\bar q^{\pm}=\frac{(S_1+2S_2)\pm\sqrt{(S_1-2S_2)^2+96a_1a_2}}{4(2-\gamma)},
\nonumber
\end{equation}
where
\begin{eqnarray}
S_1&=&(2-\gamma)(b_1-2)+3a_1,\label{S1}\\
S_2&=&(2-\gamma)(b_2-1)+4a_2.\label{S2}
\end{eqnarray}
The remaining two singular points lie in the  ${\cal BV}$ invariant set.
They are
\begin{eqnarray}
&&\left( \frac{(1-b_2)}{3a_2}, +\sqrt{\frac{(1-b_2)B_1}{3a_2(2-b_1)}},
	 \frac{3a_1(1-b_2)}{a_2(2-b_1)}, +\frac{1}{6}\sqrt{\frac{(1-b_2)B_1}{3a_2(2-b_1)}}
	 \right),\nonumber\\
&&\left( \frac{(1-b_2)}{3a_2}, -\sqrt{\frac{-(1-b_2)B_1}{3a_2(2-b_1)}},
	 \frac{3a_1(1-b_2)}{a_2(2-b_1)}, -\frac{1}{6}\sqrt{\frac{-(1-b_2)B_1}{18a_2(2-b_1)}}
	 \right),\label{BV1}
\end{eqnarray}
where $B_1$ is given by equation (\ref{B1}). 
 
The stability of each of these singular points is very difficult to determine in general.  However, one question that can be asked
is whether these anisotropic models generally isotropize; that is,
``Does there exist a stable ($t$-time) equilibrium point in the set
${\cal FRW}$?
(see subsection \ref{isotropic}).
We shall also analyze the model when there is zero anisotropic stress ($z=0$) in order to determine the effects that bulk viscous pressure may have on the models (see
subsection \ref{zero stress}).
We shall also  analyze the effect of anisotropic stress in an anisotropic 
model with zero bulk viscous pressure ($y=0$). (see subsection \ref{zero pressure}).

\subsubsection{Stability of Isotropic Singular Points}\label{isotropic}

In this subsection we are going to resolve the stability of the isotropic
singular points, that is, those singular points lying in the set ${\cal FRW}$.
We want to determine if there exists a stable ($t$-time) singular point in
the future.  In $\Omega$-time this translates to showing that there exists
a source in the set ${\cal FRW}$.

The singular point $(0,0,0,0)$ represents the Milne model, and has eigenvalues
\begin{eqnarray}
2,&2(b_2-1), \frac{1}{2}\left\{(3\gamma+b_1-4)+\sqrt{(3\gamma+b_1-4)^2
+4B_1}\right\},\nonumber\\
&\qquad\qquad\qquad
 \frac{1}{2}\left\{(3\gamma+b_1-4)-\sqrt{(3\gamma+b_1-4)^2
+4B_1}\right\},
\end{eqnarray}
where $B_1$ is given by equation (\ref{B1}).
 The bifurcation values are $b_2=1$ and $B_1=0$.  If $B_1=0$ then there exists a non-isolated line of singular points.  The stability of this point is summarized in Table \ref{tab1}.

The singular point $(1,0,y^-,0)$ represents a flat viscous fluid FRW model and has eigenvalues 
\begin{eqnarray}
&&-\frac{1}{2}\left\{(b_1+3\gamma-4)-\sqrt{(b_1+3\gamma-4)^2+4B_1}\right\},\qquad\sqrt{(b_1+3\gamma-4)^2+4B_1},\nonumber\\
&&\qquad\frac{1}{4}\left\{B_3+\sqrt{B_3^{\ 2}-8B_2}\right\},\qquad\frac{1}{4}\left\{B_3-\sqrt{B_3^{\ 2}-8B_2}\right\},
\end{eqnarray}
where
\begin{eqnarray}
B_2&=&(2b_2-3\gamma-y^-)(6-3\gamma-y^-)+24a_2,\\
B_3&=&4b_2-3(3\gamma-2)-3y^-.
\end{eqnarray}
The stability of this point is summarized in Table \ref{tab2}.

The singular point $(1,0,y^+,0)$ represents a flat viscous fluid FRW model and has eigenvalues 
\begin{eqnarray}
&&-\frac{1}{2}\left\{(b_1+3\gamma-4)+\sqrt{(b_1+3\gamma-4)^2+4B_1}\right\},\qquad-\sqrt{(b_1+3\gamma-4)^2+4B_1},\nonumber\\
&&\frac{1}{4}\left\{B_5+\sqrt{B_5^{\ 2}-8B_4}\right\},\qquad\frac{1}{4}\left\{B_5-\sqrt{B_5^{\ 2}-8B_4}\right\},
\end{eqnarray}
where
\begin{eqnarray}
B_4&=&(2b_2-3\gamma-y^+)(6-3\gamma-y^+)+24a_2,\\
B_5&=&4b_2-3(3\gamma-2)-3y^+.
\end{eqnarray}
The stability of this point is summarized in Table \ref{tab3}.

From the stability analysis of these singular points we can conclude that there exists a range of parameter values such that one of the singular points in the set ${\cal FRW}$ is a source (sink in $t$-time).
If $B_1<0$ and $b_2>1$ then the point $(0,0,0,0)$ is a source --- this result is similar to the observation of Coley and van den Hoogen \cite{Coley94a} in the Eckart theory when $m=n=1$ and $9\zeta_o-(3\gamma-2)<0$.  If $B_1>0$, $B_2>0$, and $B_3>0$ then the point
$(1,0,y^-,0)$ is a source.  However, if either of these two conditions are not satisfied then the anisotropic models will not tend to an isotropic FRW model to the future ($t$-time).  
Romano and Pav\'on \cite{Romano93,Romano94} remarked that the anisotropy dies away quickly in the anisotropic models and hence the cosmological model isotropizes,  nonetheless the cosmological model does not tend to an FRW or de Sitter model.  The same result is true here for some range of parameter values.  If $b_2<1$ and $B_1<0$ then the models all isotropize but the anisotropic stress does not tend to zero and therefore the model does not asymptotically approach an FRW model.

\subsubsection{Zero Anisotropic Stress}\label{zero stress}

In order to observe the effects of bulk viscous pressure in the model we set  $\Pi_1=\Pi_2=0$, which implies that $1/\beta_2=0$.  In the model
under consideration here this amounts to setting $z=0$ and $a_2=0$ in system (\ref{zero heat eqs}).
The resulting system is three-dimensional and has the form
\begin{mathletters}\label{sys of eqs 3}
\begin{eqnarray}
x'&=&x(3\gamma-2-2q)+y,\\
\Sigma'&=&  \Sigma(2-q),\\
y'     &=&  y(b_1-2-2q)+9a_1x,
\end{eqnarray}
\end{mathletters}
where
\begin{equation}
q     =\frac{1}{2}\left((3\gamma-2)x+y+\Sigma^2\right),
\end{equation}
and the physical phase space is
\begin{equation}
4-4x-\Sigma^2\geq 0, \qquad \text{ and }\qquad x\geq 0.
\end{equation}

In this case there exist four invariant sets of particular interest.
Similar to the previous analysis we have the set
${\cal FRW}:=\{(x,\Sigma,y)\vert \Sigma=0\}$ and ${\cal BI}:=\{(x,\Sigma,y)\vert 4-4x-\Sigma^2=0\text{ and }\Sigma\not =0\}$.  The Bianchi type V invariant set can be subdivided into two disjoint sets, ${\cal BV}^+:={\cal BI}^c\cap{\cal FRW}^c\cap\{(x,\Sigma,y)\vert\Sigma>0\}$ and ${\cal BV}^-:={\cal BI}^c\cap{\cal FRW}^c\cap\{(x,\Sigma,y)\vert\Sigma<0\}$.
Due to the symmetry in the equations (reflection through the $\Sigma=0$ plane) the qualitative behaviour in the set $\Sigma<0$ is equivalent to that in the set $\Sigma>0$; henceforth ( and without loss of generality), we shall only concern ourselves with the part of the phase space with $\Sigma\geq 0$.

The singular points of the system (\ref{sys of eqs 3})  are
\begin{equation}  (0,2,0), \quad (0,0,0),  \quad  (1,0, y^-), \quad  (1,0, y^+),  \label{singul points}
 \end{equation}
where $y^\pm$ is given by equation (\ref{y pm}).

There is only one singular point in the invariant set $\Sigma>0$.
The singular point $(0,2,0)$ is in the set ${\cal BI}$ and has eigenvalues
\begin{eqnarray}
&&-4,\quad \frac{b_1+3\gamma-12}{2}-\frac{1}{2}{\sqrt{(b_1-3\gamma)^2+36a_1}},\nonumber\\
&&\qquad \frac{b_1+3\gamma-12}{2}+\frac{1}{2}{\sqrt{(b_1-3\gamma)^2+36a_1}}.\label{3.25}
\end{eqnarray}
This singular point is either a saddle or a sink depending on the value of the parameter
\begin{equation}
B_6=(2-\gamma)(b_1-6)+3a_1.  %perhaps B_2
\end{equation}
If $B_6>0$, then the point is a saddle with a 2-dimensional stable manifold.  
If $B_6<0$, then the point is a sink, and if $B_6=0$, then the point is degenerate (discussed later).
   The solution at this singular point is a Kasner model.
The stability of this point is summarized in Table \ref{tab4}.
  
The singular point   $(0,0,0)$ has eigenvalues
\begin{eqnarray}
&& 2,
\quad
 \frac{1}{2}\left\{(b_1+3\gamma-4)+\sqrt{(b_1+3\gamma-4)^2
+4B_1}\right\}, \nonumber\\
&&\quad 
\frac{1}{2}\left\{(b_1+3\gamma-4)-\sqrt{(b_1+3\gamma-4)^2
+4B_1}\right\}.\label{000eigen}
\end{eqnarray}
This point is either a saddle or a source depending on the value of the parameter $B_1$.
If $B_1>0$, then the point is a saddle  with a 1-dimensional stable manifold.
If $B_1<0$, then the point is a source, and if $B_1=0$ the point becomes degenerate (discussed later).
The stability of this point is summarized in Table \ref{tab4}.

The singular point $(1,0,y^-)$ has eigenvalues
\begin{eqnarray}
&&-\frac{1}{2}\left\{(b_1+3\gamma-4)-\sqrt{(b_1+3\gamma-4)^2+4B_1}\right\},
\qquad
\sqrt{(b_1+3\gamma-4)^2+4B_1},\nonumber\\
&& \quad
-\frac{1}{4}\left\{(b_1+3\gamma-12)-\sqrt{(b_1-3\gamma)^2+36a_1}\right\}.\label{10y-eigen}
\end{eqnarray}
This point is either a saddle or a source depending on the value of the parameter $B_1$.
 If $B_1<0$, then the point is a saddle point with a 1-dimensional stable manifold.  
If $B_1>0$, then the point is a source, and if $B_1=0$ the point becomes degenerate (discussed later). 
The stability of this point is summarized in Table \ref{tab4}.

 The singular point $ (1,0,y^+)$ has eigenvalues
\begin{eqnarray}
&&-\frac{1}{2}\left\{(b_1+3\gamma-4)+\sqrt{(b_1+3\gamma-4)^2+4B_1}\right\},
\quad
-\sqrt{(b_1+3\gamma-4)^2+4B_1},\nonumber\\
&&\quad
-\frac{1}{4}\left\{(b_1+3\gamma-12)+\sqrt{(b_1-3\gamma)^2+36a_1}\right\}.\label{10y+eigen}
\end{eqnarray}
This singular point is either a saddle or a sink depending on the parameter $B_6$.
 If $B_6<0$, then the point is a saddle with a 2-dimensional stable manifold. 
If $B_6>0$, then the point is a sink, and if $B_6=0$ then the point is degenerate (discussed later).
The stability of this point is summarized in Table \ref{tab4}.

The bifurcations in this model occur at $B_1=0$ and $B_6=0$.  
If $B_1=0$ then there exists a line of singular points passing through the points $(0,0,0)$ and $(1,0,y^-)$.  This line can be shown to have some saddle-like properties.  In particular, if $B_1=0$ then the points $(1,0,y^-)$ and $(0,0,0)$ experience a saddle-node bifurcation.
If $B_6=0$ then the curve $y=3(2-\gamma)x$, $\Sigma^2=4-4x$ which lies in the set ${\cal BI}$ is singular. This observation is analogous to the case $\gamma=2$ in perfect fluid Bianchi type V models \cite{Coley94a}.  In particular, 
if $B_6=0$ then the points $(1,0,y^+)$ and $(0,2,0)$ experience a saddle-node bifurcation.

All information about the singular points is summarized in Table~\ref{tab4}.  It is very easily seen that if $B_6>0$ then the solutions tend to an isotropic model both to the past and to the future (in t-time), while if $B_6<0$, then solutions only tend to an isotropic model to the future.  Provided that there are no periodic or closed orbits in the set $\Sigma>0$, all models isotropize to the future ($\Omega \to -\infty$ or $t\to \infty$).  Note  the difference in the result here and the result in the previous subsection.  If there is a `non-zero' anisotropic stress then there is a range of parameter values such that models will not isotropize, and if there is `zero' anisotropic stress then all models will isotropize to the future.  Therefore we can conclude that in the truncated Israel-Stewart theory the anisotropic stress plays a dominant role in determining the future evolution of the anisotropic models. This result is contrary to  the observations in Coley and van den Hoogen \cite{Coley94a} based upon the Eckart theory where the anisotropic stress played only a minor role 
and did not determine the the future evolution of the models.

\subsubsection{Zero Bulk Viscous Pressure}\label{zero pressure}

As we have seen in subsection~\ref{isotropic}, anisotropic stress plays
a dominant role in the evolution of the anisotropic cosmological models.
To further analyze the effects of anisotropic stress on the evolution of an anisotropic model we shall set $\Pi\equiv0$, which in our model implies that $1/\beta_o=0$.   This translates into setting $y=0$ and $a_1=0$ in equations (\ref{zero heat eqs}).
In order to illustrate the possible influence anisotropic stress may have on an anisotropic model we further restrict ourselves to the set ${\cal BI}:=\{(x,\Sigma,z)\vert 4-4x-\Sigma^2=0\}$.  The resulting system is planar and lends itself easily to a complete qualitative analysis.

Consequently the system under consideration is
\begin{mathletters}\label{sys of eqs 4}
\begin{eqnarray}
\Sigma'  &=&  \frac{3}{8}(2-\gamma)\Sigma(4-\Sigma^2)-12z,\\
z'       &=&  (2b_2-3\gamma)z-\frac{3}{4}(2-\gamma)z\Sigma^2
                       +\frac{a_2}{4}\Sigma(4-\Sigma^2).
\end{eqnarray}
\end{mathletters}
The singular points are $(0,0)$, $(+2,0)$, $(-2,0)$, $(\Sigma^+,z^+)$, and
 $(\Sigma^-,z^-)$ where
\begin{equation}
\Sigma^\pm = \pm\sqrt{4+\frac{8}{3(2-\gamma)^2}B_7},\qquad\quad
z^\pm      = \frac{(2-\gamma)}{32}\Sigma^\pm(4-{\Sigma^\pm}^2),
\end{equation}
where
\begin{equation}
B_7=(2-\gamma)(b_2-3)+4a_2.
\end{equation}

The point $(0,0)$ represents a flat FRW model; the eigenvalues of this point are
\begin{equation}
\frac{1}{4}\left\{B_8\pm\sqrt{B_8^{\ 2}-48[B_7+\frac{3}{2}(2-\gamma)^2]}\right\},
\end{equation}
where
\begin{equation}
B_8=4b_2-3(3\gamma-2).
\end{equation}
If $B_7+\frac{3}{2}(2-\gamma)^2<0$, then the point $(0,0)$ is a saddle point.
If $B_7+\frac{3}{2}(2-\gamma)^2>0$, then the stability of the point $(0,0)$
depends on the parameter $B_8$.  If $B_8>0$ the point $(0,0)$ is a source and if $B_8<0$ the point $(0,0)$ is a sink.  Bifurcations of this point occur when
$B_7=-\frac{3}{2}(2-\gamma)^2$ and $B_8=0$ and are discussed later.

The points $(\pm 2,0)$ represent Kasner models.  The eigenvalues are
\begin{equation}
\frac{1}{2}\left\{(2b_2+3\gamma-12)\pm\sqrt{(2b_2+3\gamma-12)^2+24B_7}\right\}.
\end{equation}
If $B_7>0$, the points $(\pm 2,0)$ are saddle points.  If $B_7<0$, then the points $(\pm2,0)$ are sinks.  The bifurcation that occurs when $B_7=0$ is discussed later.

The points $(\Sigma^\pm,z^\pm)$ only exist when $B_7+\frac{3}{2}(2-\gamma)^2>0$.
The eigenvalues are
\begin{eqnarray}
&&\frac{1}{2}\Biggl\{\left(2b_2+3\gamma-12-\frac{5}{(2-\gamma)}B_7\right)\Biggr.
\nonumber\\
&&\qquad\qquad\pm\Biggl.
\sqrt{\left(2b_2+3\gamma-12-\frac{5}{(2-\gamma)}B_7\right)^2-24B_7\left(1+
\frac{2}{3(2-\gamma)^2}B_7\right)}\Biggr\}.
\end{eqnarray}
If $-\frac{3}{2}(2-\gamma)^2<B_7<0$, then the points $(\Sigma^\pm,z^\pm)$ are saddle points.  If $B_7>0$, then the points $(\Sigma^\pm,z^\pm)$ are sinks.  The bifurcation values $B_7=-\frac{3}{2}(2-\gamma)^2$ and $B_7=0$ are discussed later.  The stability of all singular points is summarized in Table \ref{tab5}.

Knowing the singular points and their eigenvalues only reveals the local behaviour of the system (\ref{sys of eqs 4}).  The determination of some of the global properties requires
investigating the existence or non-existence of periodic orbits and analyzing points at infinity.

The existence of periodic orbits is difficult to prove.  However, with the aid of Dulac's criterion \cite{Andronov}, we are able to prove the non-existence of periodic orbits for a range of parameter values.  Taking the divergence of the system (\ref{sys of eqs 4}) we can see that
\begin{equation}
\nabla \cdot f = \frac{1}{2}B_8-\frac{15}{8}(2-\gamma)\Sigma^2.
\end{equation}
Therefore, if $B_8<0$ then there do not exist any periodic orbits.

To analyze the points at infinity we first change to polar coordinates
$r^2=\Sigma^2+z^2$ and $\theta=\tan^{-1}(z/\Sigma)$ and then we compactify the phase space through the following transformations:
\begin{equation}
\bar r=\frac{r}{1+r},\qquad \bar \theta=\theta,\qquad \frac{d\Omega}{d\bar t}=(1-\bar r)^2,
\end{equation}
in which case the evolution equations for $\bar r$ and $\bar \theta$ become:
\begin{mathletters}
\begin{eqnarray}
\frac{d\bar r}{d\bar t} &=&      (1-\bar r)^3\bar r\left\{
         \frac{3}{2}(2-\gamma)\cos^2\bar\theta+(2b_2-3\gamma)\sin^2\bar\theta+
         (a_2-12)\cos\bar\theta\sin\bar\theta\right\}\nonumber\\
&&-(1-\bar r)\bar r^3 \frac{\cos^2\bar\theta}{4}
       \left\{\frac{3}{2}(2-\gamma)\cos^2\bar\theta+3(2-\gamma)\sin^2\bar\theta+
       a_2\cos\bar\theta\sin\bar\theta\right\},\\
\frac{d\bar \theta}{d\bar t} &=&      (1-\bar r)^2 \left\{(2b_2-\frac{3}{2}\gamma-3)\cos\bar\theta\sin\bar\theta 
           +a_2\cos^2\bar\theta +12 \sin^2\bar\theta\right\}\nonumber\\
&&-\bar r^2\frac{\cos^3\bar\theta}{4}\left\{\frac{3}{2}(2-\gamma)\sin\bar\theta
+ a_2\cos\bar\theta\right\}.
\end{eqnarray}
\end{mathletters}
The points at $r=\infty$ are mapped to the unit circle $\bar r =1$.  Hence the singular points at infinity are those points on the unit circle $\bar r =1 $ where $\frac{d\bar\theta}{d\bar t}=0$.  The singular points are thus
\begin{equation}
(1,\frac{\pi}{2}), \qquad (1,-\frac{\pi}{2}), \qquad (1,\theta^*), \quad \text{and}\quad (1,\theta^*+\pi),
\end{equation}
where
$\tan \theta^*=-2a_2/3(2-\gamma)$.
In order to determine the stability of these singular points we need to study the values of $\frac{d\bar r}{d\bar t}$ and $\frac{d\bar \theta}{d\bar t}$ in a neighborhood of each of the singular points.  We find that the points $(1,\theta^*)$ and $(1,\theta^*+\pi)$ are saddle-points while the points $(1,\pm \pi/2)$ are sources.

To obtain a complete picture of the qualitative behaviour of the model we must discuss the various bifurcations that occur.
A bifurcation occurs at $B_7=-\frac{3}{2}(2-\gamma)^2$, in which case the point $(0,0)$ undergoes a pitchfork bifurcation \cite{Wiggins} to create the two new singular points $(\Sigma^\pm, z^\pm)$ and its stability is transferred to them.  When $B_7>-\frac{3}{2}(2-\gamma)^2$, the point $(0,0)$ experiences an Andronov-Hopf bifurcation at $B_8=0$ \cite{Wiggins}.  Therefore, it can be shown that there exists a $\delta>0$ such that for every $B_8 \in (0,\delta)$ there exists a periodic orbit.  In addition, the periodic orbit is an attractor.  A third bifurcation occurs at $B_7=0$ when the points $(\Sigma^\pm,z^\pm)$ and $(\pm2,0)$ undergo a transcritical bifurcation \cite{Wiggins} in which they exchange stability.  The stability of all of the singular points (finite and infinite) is given in Table~\ref{tab5}.

\begin{figure}
{\[\epsfysize=10cm \epsfbox{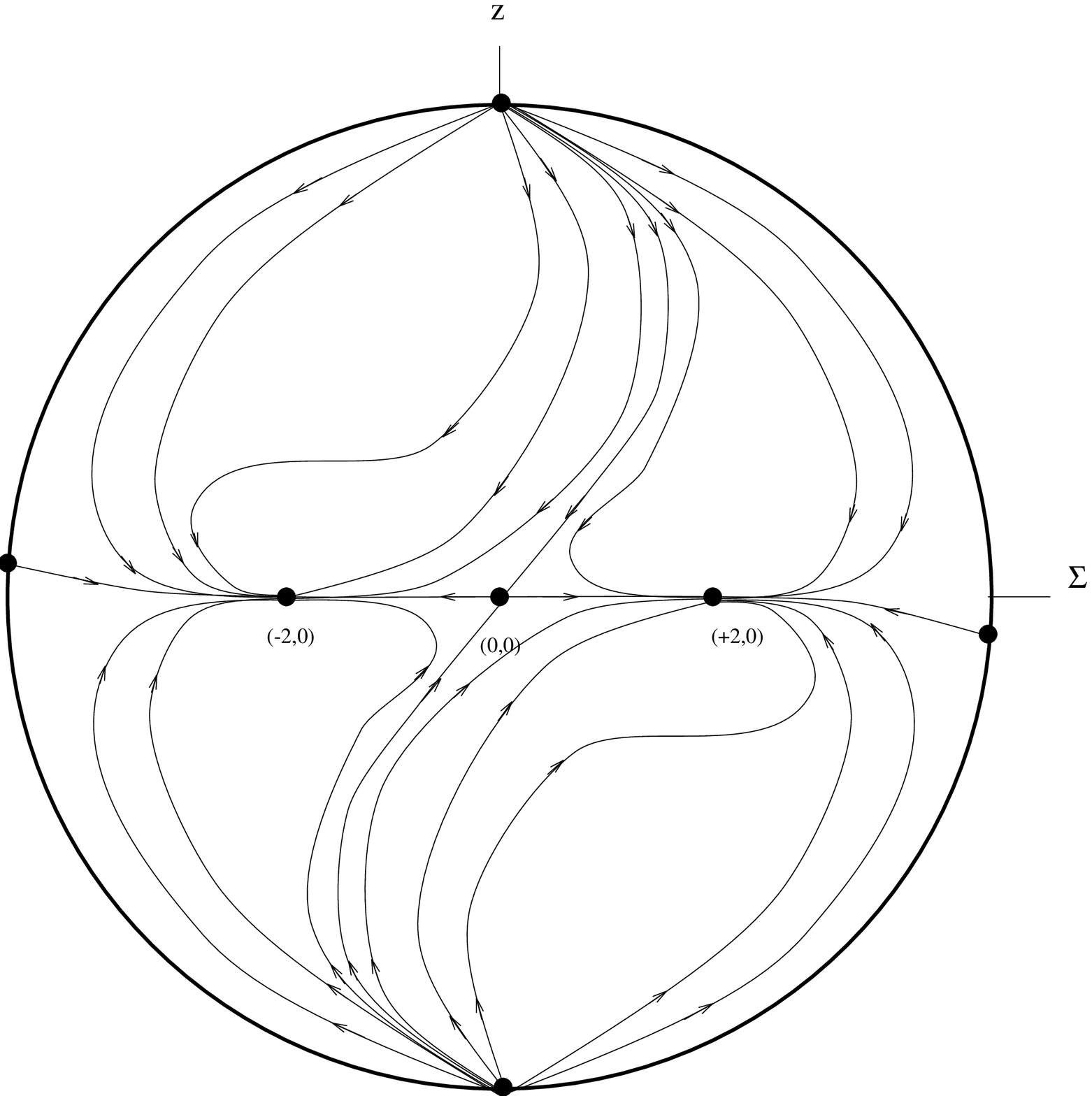}\]}
\caption{The phase portrait describes the qualitative behavior
of the anisotropic Bianchi type I  models with anisotropic stress and zero bulk viscous pressure in the case $B_7<-\frac{3}{2}(2-\gamma)^2$.  The arrows in the figure denote increasing $\Omega$-time ($\Omega\to\infty$) or decreasing $t$-time ($t\to 0^+$).}\label {fig3.1}
\end{figure}

Let us now discuss the qualitative properties of this model.
If $B_7<-\frac{3}{2}(2-\gamma)^2$, then all trajectories evolve from the singular points at $(\pm 2,0)$ representing Kasner models  to points at infinity.  These models are generally unsatisfactory since the  WEC   (which implies $\Sigma^2\leq4$ for the Bianchi I models here), is broken eventually.  However, there do exist two exceptional trajectories for which the WEC is satisfied always.  These are the trajectories that lie on the unstable manifold of the singular point $(0,0)$ which describe models that have a Kasner-like behaviour in the past (t-time) and isotropize to the future toward the point $(0,0)$.  A phase portrait of this model is given
in Figure \ref{fig3.1}.

\begin{figure}
{\[\epsfysize=10cm \epsfbox{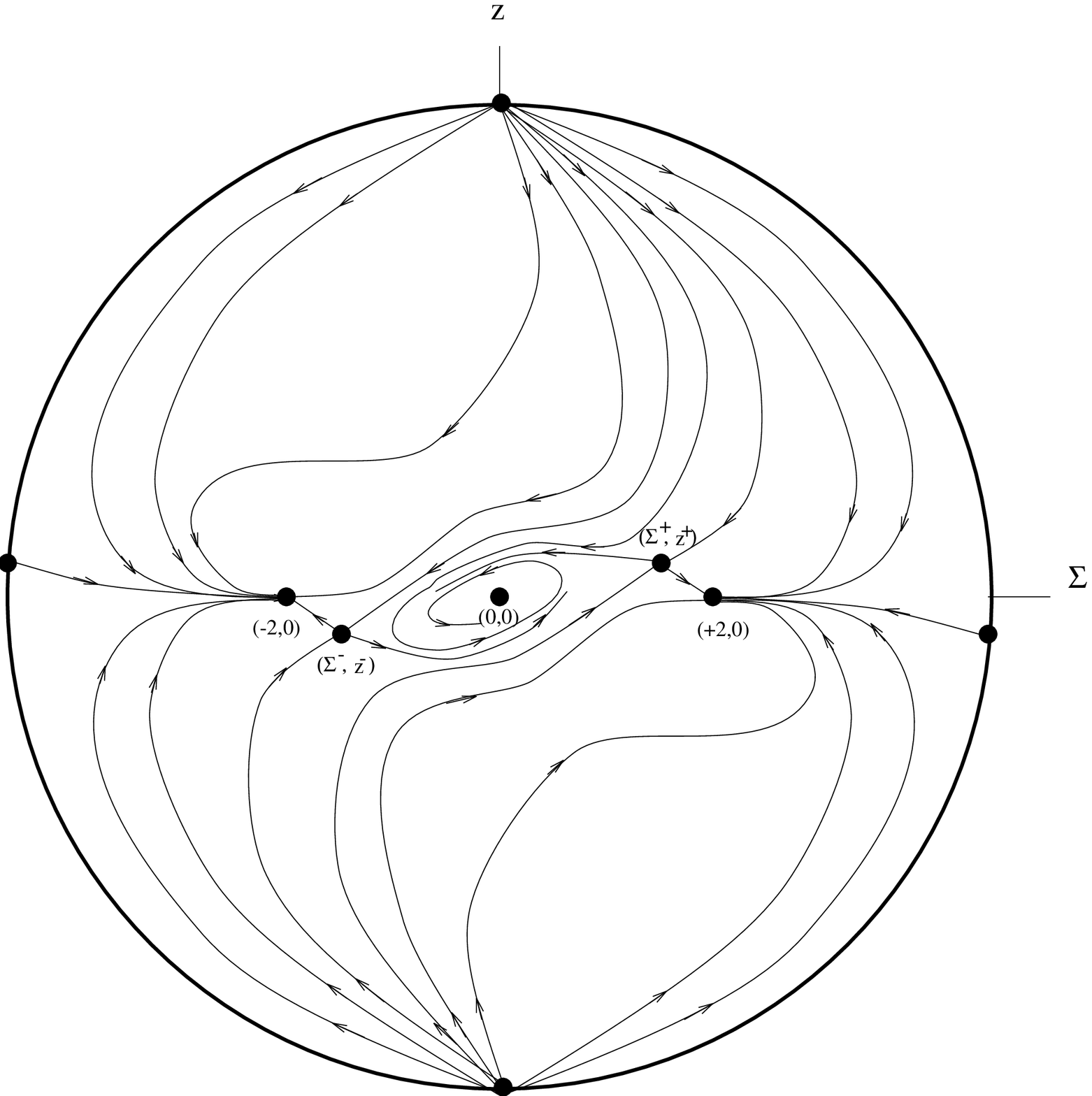}\]}
\caption{The phase portrait describes the qualitative behavior
of the anisotropic Bianchi type I  models with anisotropic stress and zero bulk viscous pressure in the case $-\frac{3}{2}(2-\gamma)^2<B_7<0$ and $B_8<0$.  The arrows in the figure denote increasing $\Omega$-time ($\Omega\to\infty$) or decreasing $t$-time ($t\to 0^+$).}\label {fig3.2}
\end{figure}

If $-\frac{3}{2}(2-\gamma)^2<B_7<0$ and $B_8<0$ then there exist two classes of generic behaviour.  One class of trajectories evolve from the isotropic singular point $(0,0)$ and evolve to points at infinity.  The second class of trajectories evolve from the singular points $(\pm2,0)$, which represent Kasner models, and evolve to points at infinity.
Both of these classes of trajectories describe models that fail to isotropize, and describe models that will eventually violate the WEC. 
If $-\frac{3}{2}(2-\gamma)^2<B_7<0$ and $B_8<0$, then there exists three sets of exceptional trajectories.  One set is the stable manifolds of the points $(\Sigma^\pm,z^\pm)$ which represent models that start at $(\Sigma^\pm,z^\pm)$ and evolve to points at infinity and hence the WEC will be violated.  
 There do exist trajectories describing models that satisfy the WEC for all time, namely the unstable manifolds of the point $(\Sigma^\pm,z^\pm)$.  One set of these trajectories  start at the singular point $(\pm2,0)$ which represent the Kasner models and evolve to the point $(\Sigma^\pm,z^\pm)$.  The other set of trajectories evolve from the isotropic singular point $(0,0)$ to the singular points $(\Sigma^\pm,z^\pm)$.  In this case there are no models which isotropize.
  A phase portrait of this model is given
in Figure \ref{fig3.2}.

\begin{figure}
{\[\epsfysize=10cm \epsfbox{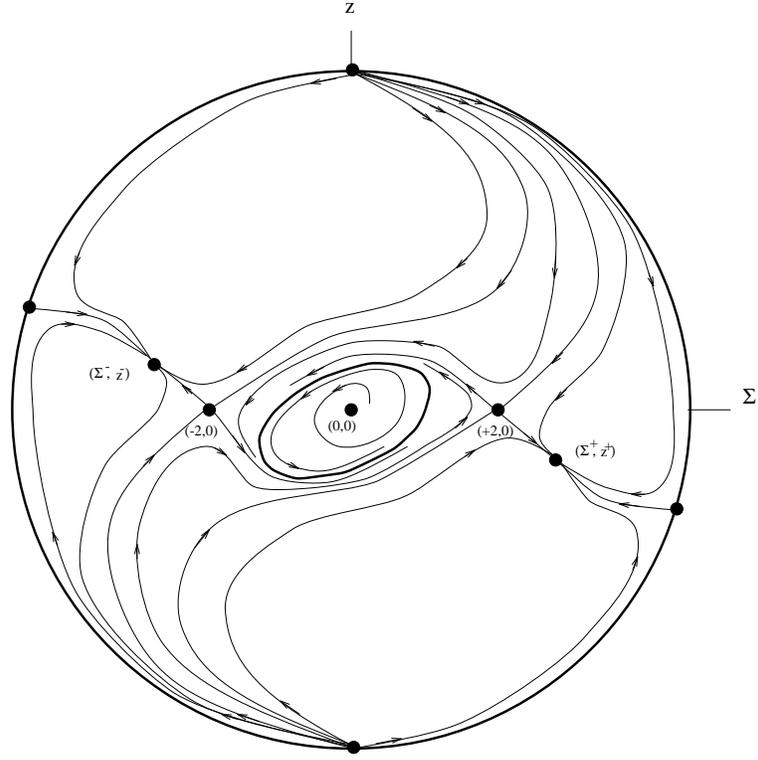}\]}
\caption{The phase portrait describes the qualitative behavior
of the anisotropic Bianchi type I  model with anisotropic stress and zero bulk viscous pressure in the case $B_7>0$ and $B_8>0$.  The closed elliptical orbit close to the center  represents the periodic orbit. The arrows in the figure denote increasing $\Omega$-time ($\Omega\to\infty$) or decreasing $t$-time ($t\to 0^+$).}\label {fig3.5}
\end{figure}

If $-\frac{3}{2}(2-\gamma)^2<B_7<0$ and $B_8>0$ then there exist three classes of models.  One class of trajectories evolves from the periodic orbit to the isotropic singular point $(0,0)$.  This class of models is interesting in that the past singularity has an oscillatory nature, that is, both the dimensionless shear $\Sigma$ and the dimensionless anisotropic stress $z$ tend to a closed periodic orbit in the past ($t$-time).  This class of trajectories also represent models that isotropize and represent   models in which the WEC is satisfied always.  The second class of trajectories are those which evolve from the periodic orbit to points at infinity.  This class of trajectories represent models that will not satisfy the WEC at some point in the future.  The third class of trajectories is the same as the second class of trajectories in the case $-\frac{3}{2}(2-\gamma)^2<B_7<0$.  Again there exist three sets of exceptional trajectories.  The stable manifolds of the points $(\Sigma^\pm,z^\pm)$ represent models that will eventually violate the WEC.  The unstable manifolds of the points $(\Sigma^\pm,z^\pm)$ represent either models that start at the Kasner-like singular-point  $(\pm2,0)$ and evolve to the point $(\Sigma^\pm,z^\pm)$ or represent models that start from the periodic orbit and evolve to the point $(\Sigma^\pm,z^\pm)$. The phase portrait in this case is very similar to that of Figure  \ref{fig3.5}.

If $B_7>0$ and $B_8<0$, then the behaviour of the trajectories is very similar to the behaviour of the trajectories in the  case $-\frac{3}{2}(2-\gamma)^2<B_7<0$ and $B_8<0$.  The difference stems from the fact that the points $(\pm2,0)$ are now saddles and the points $(\Sigma^\pm,z^\pm)$ are now sinks. The phase portrait in this case is very similar to that of Figure  \ref{fig3.2}.

If $B_7>0$ and $B_8>0$, then the behaviour of the trajectories is very similar to the behaviour of the trajectories in the case $-\frac{3}{2}(2-\gamma)^2<B_7<0$ and $B_8>0$.  The difference stems from the fact that the points $(\pm2,0)$ are now saddles and $(\Sigma^\pm,z^\pm)$ are now sinks.  A phase portrait of this model is given
in Figure \ref{fig3.5}.

In conclusion, the general behaviour of these models is unsatisfactory in that the WEC is violated eventually, except in the case $B_8>0$ and $-\frac{3}{2}(2-\gamma)^2<B_7$ where there exists a set of models (of non-zero measure) that will always satisfy the WEC.  These are the models represented by the trajectories which start at the periodic orbit and isotropize to the point $(0,0)$   to the future (t-time). 
There also exist models which satisfy the WEC always, but these are the models represented by the unstable manifolds of the saddle-points.

Clearly the anisotropic stress in the truncated-Israel-Stewart theory plays a very dominant role in the evolution of the anisotropic models.   This  is in contrast to what was found in \cite{Coley94a} using the Eckart theory  where it was found that anisotropic stress played a very minor role in determining the asymptotic behaviour.  However, if $B_8<0$, then all models are generically unsatisfactory in that the WEC will be violated.  If $B_8>0$, then there does exist a set of satisfactory models where the WEC will always be satisfied. 
It is also interesting to note briefly the existence of a periodic orbit, this type of behaviour is not seen in the Eckart models.   With the existence
of this periodic orbit, the past attractor which this periodic orbit represents, has a oscillatory character to it, in that the dimensionless shear (and therefore the dimensionless density) and the dimensionless anisotropic stress will  have an oscillatory nature.

\subsection{Non-Zero Heat Conduction}\label{III.3}

In this section we will study the effects of heat conduction on the models.  For simplicity we will assume that the anisotropic stress is zero.
Although $\Pi_{ab}\equiv0$, the bulk viscosity is still present, and hence, in a sense, we are investigating the effect heat conduction will have on the viscous models with bulk viscosity. 
Under these assumptions, the system (\ref{sys of equations}) reduces to a one-parameter family (that is, in addition to the parameters arising in the equations of state) of three-dimensional systems \cite{Coley94a}.   We label this new parameter $C$, which is a function of the integration constant that appears when equations (\ref{Sigma1}) and (\ref{Sigma2}) are integrated. The parameter $C=(k+1)/\sqrt{k^2-k+1}$ is bounded between $-1\leq C \leq 2$ and is analogous to the parameter used in \cite{Coley94a}.  The value $C=0$ corresponds to the case in which there is zero heat conduction, and $C=2$ corresponds to the case in which the model is locally rotationally symmetric (LRS).  
We define a new shear variable 
\begin{equation}
\Sigma\equiv\frac{\Sigma_1+\Sigma_2}{\sqrt{3}C},
\end{equation}
whence the system becomes\begin{mathletters}\label{heat eqs}
\begin{eqnarray}
x'&=&x(3\gamma-2-2q)+y-C\frac{\Sigma}{4}(4-4x-\Sigma^2),\\
\Sigma'&=&  \Sigma(2-q),\\
y'     &=&  y(b_1-2-2q)+9a_1 x+\sqrt{3}Cc_1\Sigma(4-4x-\Sigma^2),
 \end{eqnarray}
\end{mathletters}
where
\begin{equation}
q     =\frac{1}{2}\left((3\gamma-2)x+y+\Sigma^2\right),
\end{equation}
and the physical phase space is
\begin{equation}
4-4x-\Sigma^2\geq 0 \qquad\text{ and }\qquad x\geq 0.\label{phase3}
\end{equation}

There exists three   physically interesting invariant sets in
the phase space of the system, namely, 
${\cal FRW}:=\{(x,\Sigma,y)\vert 
\Sigma=0\}$, ${\cal BI}:=\{(x,\Sigma,y) \vert 4-4x-\Sigma^2=0, \text{ and }  \Sigma \not = 0\}$, and ${\cal BV}:= {\cal BI}^c \cap {\cal FRW}^c$ (where subscript $c$ denotes the 
complement) which represents Bianchi type V models.
 As before, the set ${\cal FRW}$ represents the spatially homogeneous
and isotropic negative and flat curvature FRW models and the set ${\cal BI}$ represents the Bianchi type I models.  

There are  six
different singular points of the system.
The singular points lying in the set ${\cal FRW}$ are:
\begin{equation}
(0,0,0),\qquad (1,0,y^-),\qquad(1,0,y^+),
\end{equation}
where $y^\pm$ is given by equation (\ref{y pm}).
Also, if $B_1=0$ then there is a non-isolated singular line that passes through the points $(0,0,0)$ and $(1,0,y^-)$, where $B_1$ is given by equation (\ref{B1}).
  These points represent open ($x=0$) and flat ($x=1$) FRW models.  The eigenvalues of the linearization in a neighborhood of each of the isotropic singular points are the same as in the case with zero heat conduction  
(see equations (\ref{000eigen}), (\ref{10y-eigen}), and (\ref{10y+eigen}) for the eigenvalues of $(0,0,0)$, $(1,0,y^-)$, and $(1,0,y^+)$, respectively, and the appropriate parts of Table \ref{tab4}).

The singular points located in the set ${\cal BI}$ are
\begin{equation}
(0,-2,0),\qquad \text{ and} \qquad (0,+2,0).
\end{equation}
The eigenvalues of the linearization about the point $(0,-2,0)$ are similar to those in the case with zero heat conduction [see equation (\ref{3.25})]; indeed, only the first eigenvalue is different, namely, instead of $\lambda_1=-4$, we now have $\lambda_1=-4-2C$, which is very easily seen to be negative definite.  Therefore the stability of the point $(0,-2,0)$ is the same as in the previous case with zero heat conduction.
Similarly, for the eigenvalues of the linearization about the point $(0,+2,0)$,  only the first eigenvalue is different, namely, instead of $\lambda_1=-4$ we now have $\lambda_1=-4+2C$, which is   negative definite for $C\not = 2$.  Therefore, if $C\not = 2$, the stability of the point $(0,+2,0)$ is the same as in the case with zero heat conduction.  The case $C=2$   is discussed below.

The sixth singular point is $(\bar x, \bar \Sigma, \bar y)$, where
\begin{eqnarray}
\bar x &=& \frac{4(C^2-4)(b_1-6+4\sqrt{3}c_1)}{C^2[16\sqrt{3}c_1+(3\gamma-2)(b_1-6)-9a_1]},\nonumber\\
\bar \Sigma&=&\frac{4}{C},\nonumber\\
\bar y&=&\frac{12(C^2-4)[4\sqrt{3}c_1(2-\gamma)-3a_1]}{C^2[16\sqrt{3}c_1+(3\gamma-2)(b_1-6)-9a_1]}.
\end{eqnarray}
However this last singular point lies outside the region of phase space defined by equations (\ref{phase3}) for $C\not=2$. 

 If $C=2$, then a `transcritical' bifurcation occurs, the points $(\bar x, \bar \Sigma, \bar y)$ and $(0,+2,0)$ coalesce and become a single point.  The stability of this point cannot be determined via linearization.     If $(3\gamma-2)(6-b_1)+9a_1-16\sqrt{3}c_1 = 0$, then there is a line of
singular points $y+(3\gamma-2)x=0$, $\Sigma=2$.  The stability of the singular point is very difficult to determine analytically (even with the use of center manifold theory \cite{Wiggins}).  However, numerical experiments in addition to some analysis show that the singular point has some of the same behaviour as in the case with $C\not = 2$ (e.g., if $B_6>0$ the point is a saddle and if $B_6<0$ then the point has both saddle-like and sink-like behaviour).

The stability of the singular points $(0,0,0)$, $(1,0,y^-)$, $(1,0,y^+)$, $(0,-2,0)$, and $(0,+2,0)$ are the same as in the case with zero heat conduction.  The heat conduction does not determine the stability of the singular points that lie in the physical phase space (\ref{phase3}) 
but does play a role in determining their eigendirections.   
This is similar to the situation in which the bulk viscous pressure is absent whence the model reduces to one that was analyzed in Coley and van den Hoogen \cite{Coley94a}; there the addition of heat conduction did not change the stability of the singular points  but did allow the models to violate the WEC by rotating the principal eigendirections.

%0000000000000000000000000000000000000000000000000000
%0000000000000000      CONCLUSION   000000000000000000
%0000000000000000000000000000000000000000000000000000

\section{Conclusion}\label{IV}

This work improves over previous work \cite{Coley94a,Abolghasem93,Coley92,Burd94} on viscous cosmology using the non-causal first-order thermodynamics of Eckart \cite{Eckart} in that a causal theory of irreversible thermodynamics has been utilized.  Also, this work enhances the analysis of anisotropic viscous cosmologies in \cite{Belinskii80,Romano93,Romano94} because more than just the isotropic singular points have been analyzed.  The present work also generalizes the analysis of causal viscous FRW models in \cite{Coley95}.  

Again we have seen that the singular points of the dynamical system describing the evolution of an anisotropic viscous fluid cosmological model are, in general, self-similar \cite{Coley94b}.  In the case in which $m=n=1$, $r_i=1$ and $p_i=-1$ ($ i =1,  2$) all singular points are self-similar except in the case in which $\gamma=3\zeta_o$ when there exists a singular point that represents a de Sitter model which is not self-similar.

We have found that in the case of   zero heat conduction the anisotropic models need not isotropize (that is, there exists a range of parameter values and initial conditions such that the models will not isotropize).  The parameter $b_2$, which is the parameter related to the relaxation time of the anisotropic stress, plays a major role in determining the stability of the isotropic models.  In the special case of zero  anisotropic stress we have shown that all models isotropize.
The addition of anisotropic stress  on an anisotropic Bianchi type I model reveals some of the effects that anisotropic stress has on an anisotropic model. For instance, anisotropic stress generically causes models to increase their anisotropy and eventually violate the weak energy condition.  Anisotropic stress in some instances causes the creation of a periodic orbit.  This periodic orbit represents a past attractor in which the dimensionless quantities $\rho/\theta^2$ and $\sigma/\theta$ are approximately periodic.    It is interesting to note that it is only when this periodic orbit  is present that there exist trajectories (in the interior) which represent models that will isotropize and satisfy the weak energy condition.   

The models with heat conduction analyzed here had no anisotropic stress but did have bulk viscosity.  Consequently, we have investigated whether any qualitative changes arise from the inclusion of heat conduction.  From our analysis
the addition of heat conduction in the model did not change the stability of the singular points, an hence  the asymptotic states of the models.    However,  the inclusion of heat conduction did affect the dynamics in a neighborhood of each of the singular points since the eigendirections changed.

In this work  we have employed
  the truncated Israel-Stewart theory which is a causal and stable second order relativistic theory of irreversible thermodynamics.   It is possible that the truncated theory is applicable in the very early universe.
However, it is   known that such a truncated theory could result in some pathological behaviour, (e.g., in the temperature \cite{Maartens94}).
Hence this work  should be considered as a first step in the analysis of the full   theory of Israel-Stewart-Hiscock  \cite{Hiscock91,Zakari93,IsraelStewart79,Pavon82}. The present paper and \cite{Coley95} provide a firm foundation for the analysis of viscous cosmological models using the full theory which we hope to present in the future.

%0000000000000000000000000000000000000000000000000000 
%0000000000      ACKNOWLEDGEMENTS       0000000000000
%0000000000000000000000000000000000000000000000000000 

\acknowledgements

This research was funded by the Natural Sciences and Engineering Research
Council of Canada and a Killam scholarship awarded to RVDH. We would also 
like to thank Diego Pavon for pointing out an error in an earlier version
of this paper.

%0000000000000000000000000000000000000000000000000000 
%0000000000           REFERENCES     0000000000000000
%0000000000000000000000000000000000000000000000000000 

%\bibliographystyle{prsty}

%\bibliography{../vis}

\begin{thebibliography}{10}

\bibitem{Coley95}
A.~A. Coley and R.~J. van~den Hoogen, Class. Quantum Grav. {\bf (accepted)},
  (1995).

\bibitem{Coley94a}
A.~A. Coley and R.~J. van~den Hoogen, J. Math. Phys. {\bf 35},  4117  (1994).

\bibitem{Abolghasem93}
G. Abolghasem and A.~A. Coley, Int. J. Theor. Phys. {\bf 33},  695  (1994).

\bibitem{Coley92}
A. Coley and K. Dunn, J. Math. Phys. {\bf 33},  1772  (1992).

\bibitem{Burd94}
A. Burd and A. Coley, Class. Quantum Grav. {\bf 11},  83  (1994).

\bibitem{IsraelStewart79}
W. Israel and J.~M. Stewart, Ann. Phys. {\bf 118},  341  (1979).

\bibitem{Belinskii80}
V.~A. Belinskii, E.~S. Nikomarov, and I.~M. Khalatnikov, Sov. Phys. JETP {\bf
  50},  213  (1979).

\bibitem{Romano93}
V. Romano and D. Pav{\'o}n, Phys. Rev. D {\bf 47},  1396  (1993).

\bibitem{Romano94}
V. Romano and D. Pav{\'o}n, Phys. Rev. D {\bf 50},  2572  (1994).

\bibitem{Coley94b}
A.~A. Coley and R.~J. van~den Hoogen,  in {\em {D}eterministic {C}haos in
  {G}eneral {R}elativity {\rm edited by D. Hobill, A. Burd, and A. Coley}}
  ({N}{A}{T}{O} {A}{S}{I} 332{B}, {P}lenum, {N}ew {Y}ork, 1994).

\bibitem{Coley90b}
A.~A. Coley, J. Math. Phys. {\bf 31},  1698  (1990).

\bibitem{Coley90a}
A.~A. Coley, Gen. Rel. Grav. {\bf 22},  3  (1990).

\bibitem{MacCallum73}
M.~A.~H. MacCallum,  in {\em {C}arg\`ese {L}ectures in {P}hysics, {\rm edited
  by E. Schatzman}} ({G}ordon and {B}reach, {N}ew {Y}ork, 1973).

\bibitem{Bluman}
G.~W. Bluman and S. Kumei, {\em {S}ymmetries and {D}ifferential {E}quations}
  ({S}pringer-{V}erlag, {N}ew {Y}ork, 1989).

\bibitem{HawkingEllis}
S.~W. Hawking and G.~F.~R. Ellis, {\em {T}he large scale structure of
  space-time} ({C}ambridge Univ. Press, {C}ambridge, 1973).

\bibitem{Kolassis88}
C.~A. Kolassis, N.~O. Santos, and D. Tsoubelis, Class. Quantum Grav. {\bf 5},
  1329  (1988).

\bibitem{Zakari93}
M. Zakari and D. Jou, Phys. Rev. D {\bf 48},  1597  (1993).

\bibitem{Hirsch}
M.~H. Hirsch and S. Smale, {\em {D}ifferential {E}quations, {D}ynamical
  {S}ystems, and {L}inear {A}lgebra} ({A}cademic {P}ress, {N}ew {Y}ork, 1974).

\bibitem{Sansone}
G. Sansone and R. Conti, {\em {N}on-{L}inear {D}ifferential {E}quations}
  ({P}ergamon {P}ress, {N}ew {Y}ork, 1964).

\bibitem{Wiggins}
S. Wiggins, {\em {I}ntroduction to {A}pplied {N}onlinear {D}ynamical {S}ystems
  and {C}haos} ({S}pringer-{V}erlag, {N}ew {Y}ork, 1990).

\bibitem{Andronov}
A.~A. Andronov, E.~A. Leontovich, I.~I. Gordon, and A.~G. Maier, {\em
  {Q}ualitative {T}heory of {S}econd-{O}rder {D}ynamic {S}ystems} ({W}iley,
  {N}ew {Y}ork, 1973).

\bibitem{Eckart}
C. Eckart, Phys. Rev. {\bf 58},  919  (1940).

\bibitem{Maartens94}
R. Maartens, 1994 (unpublished).

\bibitem{Hiscock91}
W.~A. Hiscock and J. Salmonson, Phys. Rev. D {\bf 43},  3249  (1991).

\bibitem{Pavon82}
D. Pav{\'o}n, D. Jou, and J. Casas-V{\'a}zquez, Ann. Inst. H. Poincar{\'e} A
  {\bf 36},  79  (1982).

\end{thebibliography}

%0000000000000000000000000000000000000000000000000000 
%0000000000           Appendix       0000000000000000
%0000000000000000000000000000000000000000000000000000 

\appendix
\section{Energy Conditions}\label{appendixA}

For an imperfect fluid the energy conditions can be formulated with respect to the eigenvalues of the energy momentum tensor \cite{Kolassis88}.  The weak energy condition (WEC) states that $T_{ab}W^aW^b\geq0$ for any timelike vector $W^a$ \cite{HawkingEllis}.  In the models under investigation the WEC, written in  dimensionless variables,   becomes 
\begin{eqnarray}
3(2-\gamma) x-y-2\sqrt{3}(z_1+z_2)+9\Delta&\geq&0,\nonumber\\
3\gamma x+y+2\sqrt{3}(z_1-5z_2)+9\Delta&\geq&0,\nonumber\\
3\gamma x+y+2\sqrt{3}(z_2-5z_1)+9\Delta&\geq&0,
\end{eqnarray}
where
\begin{equation}
\Delta\equiv \frac{1}{18}\sqrt{[6\gamma+2y+6\sqrt{3}(z_1+z_2)]^2-3(\Sigma_1+\Sigma_2)^2(4-4x-\Sigma^2)}.\label{delta}
\end{equation}
We note that the eigenvalues of the energy momentum tensor must be real \cite{Kolassis88} and therefore the quantity under the square root sign in (\ref{delta}) must be positive.

The dominant energy condition (DEC) states that for every timelike $W^a$, $T_{ab}W^aW^b\geq0$ and $T^{a}_{\ b}W^b$ is non-spacelike \cite{HawkingEllis}.  
In the models under investigation the DEC    becomes  
\begin{eqnarray}
0&\leq&3(2-\gamma) x-y-2\sqrt{3}(z_1+z_2)\nonumber\\
0&\leq&3\gamma x+y+2\sqrt{3}(z_1-5z_2) +9\Delta\leq 6(2-\gamma)x -2y-4\sqrt{3}(z_1+z_2) +18\Delta,\nonumber\\
0&\leq&3\gamma x+y+2\sqrt{3}(z_2-5z_1) +9\Delta\leq 6(2-\gamma)x -2y-4\sqrt{3}(z_1+z_2) +18\Delta.
\end{eqnarray}

The strong energy condition (SEC) states that  $T_{ab}W^aW^b-\frac{1}{2}T_a^{\;a}W^bW_b\geq0$  for any timelike vector $W^a$ \cite{HawkingEllis}.  
In the models under investigation the SEC    becomes  
\begin{equation}
\text {WEC and}\quad 6(\gamma-1) x+2y-2\sqrt{3}(z_1+z_2)+9\Delta\geq 0.
\end{equation}

%0000000000000000000000000000000000000000000000000000 
%0000000000000000000      TABLES    00000000000000000
%0000000000000000000000000000000000000000000000000000 

%0000000000000000000000000000000000000000000000000000 
%0000000000000000000      TABLE 1   00000000000000000
%0000000000000000000000000000000000000000000000000000 
 
\mediumtext
\begin{table}[p]
\caption{Stability of the singular point $(0,0,0,0)$ where $\dim(W^s)$ 
is the dimension of the stable manifold with respect to $\Omega$-time.}
\label{tab1}
\begin{tabular}{llll}
$\sgn(B_1)$ & $\sgn(b_2-1)$ & type & $\dim(W^s)$ \\
\tableline
$+$ & $+$ & Saddle & 1 \\
$+$ & $-$ & Saddle & 2 \\
$-$ & $+$ & Source & 0 \\
$-$ & $-$ & Saddle & 1 \\
\end{tabular}
%\tablenotetext [1] {$B_1=(3\gamma-2)(2-b_1)+9a_1$.}
\end{table}

%0000000000000000000000000000000000000000000000000000 
%0000000000000000000      TABLE 2   00000000000000000
%0000000000000000000000000000000000000000000000000000 
 
\begin{table}[p]
\caption{Stability of the singular point $(1,0,y^-,0)$ where $\dim(W^s)$ 
is the dimension of the stable manifold with respect to $\Omega$-time.}
\label{tab2}
\begin{tabular}{lllll}
$\sgn(B_1)$ & $\sgn(B_2)$  & $\sgn(B_3)$  & type & $\dim(W^s)$ \\
\tableline
$+$ & $-$ &     & Saddle & 1 \\
$+$ & $+$ & $+$ & Source & 0 \\
$-$ & $-$ &     & Saddle & 2 \\
$-$ & $+$ & $+$ & Saddle & 1 \\
$-$ & $+$ & $-$ & Saddle & 3 \\
\end{tabular}
%\tablenotetext [1] {$B_1=(3\gamma-2)(2-b_1)+9a_1$.}
%\tablenotetext [2] {$B_2=(2b_2-3\gamma-y^-)(6-3\gamma-y^-)+24a_2$.}
%\tablenotetext [3] {$B_3=4b_2-3(3\gamma-2)-3y^-$.}
\end{table}

%0000000000000000000000000000000000000000000000000000 
%0000000000000000000      TABLE 3   00000000000000000
%0000000000000000000000000000000000000000000000000000 
 
\begin{table}[p]
\caption{Stability of the singular point $(1,0,y^+,0)$ where $\dim(W^s)$ 
is the dimension of the stable manifold with respect to $\Omega$-time.}
\label{tab3}
\begin{tabular}{llll}
 $\sgn(B_4)$ & $\sgn(B_5)$ & type & $\dim(W^s)$ \\
\tableline
$-$ &     & Saddle & 3 \\
$+$ & $+$ & Saddle & 2 \\
$+$ & $-$ & Sink    & 4 \\
\end{tabular}
%\tablenotetext [1] {$B_4=(2b_2-3\gamma-y^+)(6-3\gamma-y^+)+24a_2$.}
%\tablenotetext [2] {$B_5=4b_2-3(3\gamma-2)-3y^+$.}
\end{table}

%0000000000000000000000000000000000000000000000000000 
%0000000000000000000      TABLE 4   00000000000000000
%0000000000000000000000000000000000000000000000000000 
 
\widetext
\begin{table}[p]
\caption{Stability of the singular points with respect to $\Omega$-time (both with and witout   heat conduction).}\label{tab4}
 \begin{tabular}{llcccc}
$\sgn(B_6)$  & $\sgn(B_1)$  & $(0,0,0)$ & $(1,0,y^-)$  & $(1,0,y^+)$  & $(0,2,0)$ \\ \tableline
% 000000000000000000000000000000000000000000000000000
$+$& $-$ & 
source & 
saddle &
sink &
saddle \\
% 000000000000000000000000000000000000000000000000000
$+$ & $+$ & 
saddle & 
source &
sink &
saddle \\
% 000000000000000000000000000000000000000000000000000
$-$&$-$ & 
source & 
saddle &
saddle &
sink \\
 % 000000000000000000000000000000000000000000000000000
$-$& $+$ & 
saddle & 
source &
saddle &
sink \\
\end{tabular}
% \tablenotetext [1] {$B_6=(2-\gamma)(b_1-6)+3a_1$.} 
% \tablenotetext [2] {$B_1=(3\gamma-2)(2-b_1)+9a_1$.}
% \tablenotetext [3] {$y^-=\left({b_1-3\gamma}-{\sqrt{(b_1-3\gamma)^2 + 36 
% a_1}}\right)/{2}$.} 
% \tablenotetext [4] {$y^+=\left({b_1-3\gamma}+{\sqrt{(b_1-3\gamma)^2 + 36
% a_1}}\right)/{2}$.}
\end{table}

%0000000000000000000000000000000000000000000000000000 
%0000000000000000000      TABLE 5   00000000000000000
%0000000000000000000000000000000000000000000000000000 

\begin{table}[p]
\caption{Stability of the singular points with respect to $\Omega$-time, at both finite and infinite values, for the Bianchi type I anisotropic model with anisotropic stress and zero viscous pressure.}
\label{tab5}
\begin{tabular}{lllllccc}
$B_7$ & $\sgn(B_8) $ & $(0,0)$ & $(\pm2,0)$ &  $(\Sigma^\pm,z^\pm)$ & $(0,\pm\infty)$\tablenotemark [1] &  $(\pm\infty,\mp\infty)$\tablenotemark [2] & periodic orbit \\
\tableline
$B_7<\frac{-3}{2}(2-\gamma)^2 $  &     & saddle & sink   &        & source & saddle & \\
$\frac{-3}{2}(2-\gamma)^2<B_7<0$ & $-$ & sink   & sink   & saddle & source & saddle & \\
$\frac{-3}{2}(2-\gamma)^2<B_7<0$ & $+$ & source & sink   & saddle & source & saddle & sink\\
$B_7>0$                          & $-$ & sink   & saddle & sink   & source & saddle & \\
$B_7>0$                          & $+$ & source & saddle & sink   & source & saddle & sink \\
 \end{tabular}
%\tablenotetext [1] {$B_7=(b_2-3)(2-\gamma)+4a_2$.} 
%\tablenotetext [2] {$B_1=4b_2-3(3\gamma-2)$.}
\tablenotetext [1] {These are the points at infinity corresponding to $(\bar r =1, \bar\theta=\pm \pi/2)$.}
\tablenotetext [2] {These are the points at infinity corresponding to $(\bar r =1, \bar\theta=\theta^*)$ and $(\bar r =1, \bar\theta=\theta^*+\pi)$.}
\end{table}

\vfill\eject
 
%0000000000000000000000000000000000000000000000000000 
%0000000000           FIGURES        0000000000000000
%0000000000000000000000000000000000000000000000000000 

\input epsf

\vfill\eject

\vfill\eject

\end{document}